# Room-Temperature Silicon Band-Edge Photoluminescence Enhanced by Spin-Coated Sol-Gel Films

S. Abedrabbo,[a,b,]* B. Lahlouh,[a] S. Shet [c] and A.T. Fiory [b]

[a] *Department of Physics, University of Jordan, Amman 11942, Jordan*
[b] *Department of Physics, New Jersey Institute of Technology, Newark, NJ 07901, USA*
[c] *National Renewable Energy Laboratory, Golden CO 80401, USA*

**Abstract**

Photoluminescence is observed at room temperature from phonon-assisted band-to-band emission in Si (1.067 eV peak) using unpatterned bulk p-type silicon wafer samples that were spin-coated with Er-doped (6 at. %) silica-gel films (0.13 μm) and vacuum annealed; the strongest emission was obtained at ~700 °C. Comparative study of annealing behavior indicates two-orders of magnitude efficiency enhancement. Emission from $Er^{+3}$ ions in the silica film is used to gauge relative emission strengths. Mechanisms for inducing emission from silicon utilizing stresses in sol-gel-films are discussed.

**Graphical Abstract**

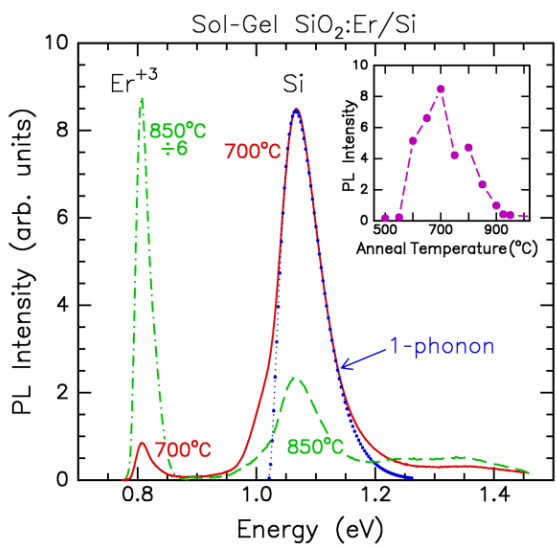

* Corresponding author: Email: sxa0215@yahoo.com

Producing light emission from silicon at room temperature generally entails selecting structures and materials to circumvent inherent disadvantages of the indirect Si band gap (e.g. typical $10^{-4}$ quantum efficiency at 300 K [1], see also [2]) [3]. Previous methods include selection of float-zone silicon with high minority carrier lifetime [4], nano structured p-n junctions [5], oxidized porous silicon [6], Si nano crystals embedded in sub-oxide [7], ultra-thin Si crystal (silicon on insulator or SOI) [8,9], nano patterned SOI [10], and p-n junctions formed by ion-implantation [11-14], diffusion doping [14], or amorphous-Si hetero structures [15]. This work reports enhancement of near-band-edge 1.067-eV (1162-nm) emission at room temperature by spin-coating and annealing sol-gel silica films on bulk silicon without employing any of these silicon-material manipulations. Enhanced Si emission is shown to be correlated with inhomogeneous film stresses.

The silica films were doped with erbium for the purpose of producing IR (infrared) emission using common silicon wafers as substrates. Consequently, photoluminescence (PL) spectra show the expected optical activity in the 1535-nm band (~0.807 eV) from $Er^{+3}$ ions in the silica film (intra 4-f band $^4I_{13/2} \to {}^4I_{15/2}$ transitions), in addition to emission from the Si substrate (1.067 eV peak). Room-temperature emission from Si in association with sol-gel films (whether Er doped or not) appears to be unreported in prior work.

Samples for this study were made from 2.5-cm pieces cut from 150-mm Czochralski (CZ) Si wafers (Semiconductor Wafer, Inc.) of specifications: boron doped p-type, 3 – 30 Ωcm, ⟨100⟩ orientation, single-side polished, 675 ± 25 μm thickness (minority carrier lifetime in CZ Si is typically rather short, ~ 50 μs). The silicon was cleaned with an RCA process. Prior to mounting in a spin coater, the Si substrates were cleaned with acetone, methanol, deionized water, and blown dry. The spin coating solution was prepared using a sol-gel recipe: 0.5 g high-purity $Er_2O_3$ powder was dissolved in a solution of 4 ml ethanol, 4 ml acetic acid, and 1.6 ml deionized water and stirred at 45 °C for 3 hours; 2 ml $Si(OC_2H_5)_4$ (TEOS) was added and stirred at 80 °C for 10 min. The resulting sol was applied through a 0.45-μm syringe filter onto Si in a spin-coater (1200 rpm, 30 sec). The samples were then oven dried in air at 120 °C for 30 minutes and subsequently annealed in a horizontal vacuum furnace (2 Pa, 1 hour) at temperatures $T_a$ ranging from 500 to 950 °C. Photoluminescence spectra were recorded by a model Fluorolog-3 spectrofluorometer (Horiba Jobin Yvon) with a double excitation monochromator (Xe lamp source), single emission monochromator, and cooled InGaAs photodiode (Hamamatsu), in the wavelength range 850 – 1600 nm (data converted to energy: 0.775 – 1.459 eV). The excitation wavelength (522 nm) was determined for producing maximum $Er^{+3}$ emissions.

Photoluminescence spectra are shown in Fig. 1 for two samples, A and B, that were annealed at $T_a$ of 700 °C and 850 °C, corresponding to solid and broken curves, respectively. The data show two major peaks at 0.807 and 1.067 eV. The 1.067-eV peak is strongest for annealing at 700 °C (sample A), where it is enhanced by 50× when compared to unannealed, air-dried films. This emission is associated with the Si substrate (as verified below). Silica film thickness of 0.13 μm (25% shrinkage upon annealing, as is typical [16]) and refractive index of 1.471 at 630 nm was measured for sample A. The 0.807-eV peak is strongest for annealing at 850 °C (sample B, left dot-dash curve, scaled by factor 1/6), and is associated with emission from $Er^{+3}$ in the silica film. Weaker emissions in the region 1.25 – 1.35 eV are attributed to $^4I_{11/2} \to {}^4I_{15/2}$ transitions in $Er^{+3}$. The annealing temperature for sample B corresponds to maximum 0.807-eV $Er^{+3}$ emission and is near that used for similar erbium-doped sol-gel silica films (~900 °C [17]). On comparing the two emission signals, the peak emission from the Si (sample A at 700 °C) is about 16% of the peak emission from $Er^{+3}$ (sample B at 850 °C); from areas under the respective spectra, the integrated signal from the Si is 50 % of that from $Er^{+3}$.

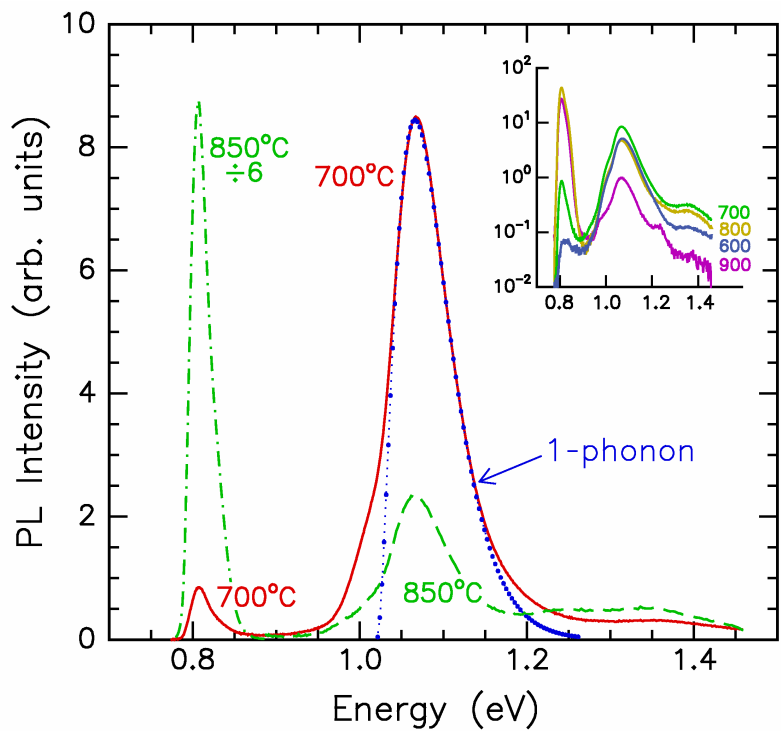

**Figure 1.** Photoluminescence intensity spectra from samples of Er-doped sol-gel films on Si annealed at temperatures of 700 °C for sample A, solid curve; and 850 °C for sample B, broken curves (left dot-dash portion divided by 6). Dotted curve shows Si one-phonon component for sample A. Inset: PL spectra (semi-log scale) at indicated annealing temperatures (°C).

Included in Fig. 1 (dotted curve) is a one-phonon (i.e. dominant phonon) model for emission of photons at energy E by phonon-assisted recombination of free electrons and holes in silicon, represented by the expression, $PL(E) \propto (E-E_0)^2 \exp[-(E-E_0)/k_BT]$ (uncorrelated pair model, see e.g. [15]); it is calculated with threshold energy $E_0 = 1.020$ eV and thermal energy $k_BT = 0.023$ eV. Since the dominant phonons are the momentum-conserving transverse optical phonons of energy $E_{ph} = 0.0578$ eV, the effective band gap, neglecting possible corrections for exciton binding or trapping, is $E_G = E_0 + E_{ph} = 1.078$ eV, which is about 42 meV below the intrinsic Si band gap (1.12 eV). Although this is an approximate examination of the data (weaker phonon components, e.g. features appearing below 1.03 eV, were not included because they commingle with $Er^{+3}$ emissions), the reasonable overlap for much of the PL signal peaking at 1.067 eV indicates that this emission most likely originates from the silicon substrate and not from the film. Figure 1, inset, presents PL spectra on semi-log scales, showing similarity in exponential form ($E>E_0$) at the various $T_a$. The X-ray diffraction, not shown, of sample A shows no evidence of polycrystalline Si (e.g. absence of Si nano crystals within the deposited film [3]).

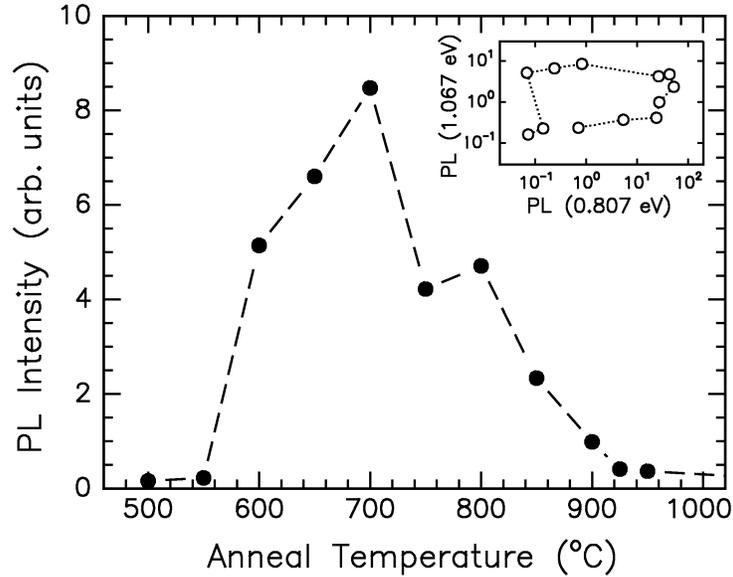

**Figure 2.** Peak photoluminescence intensities (~1.067 eV) from samples of Er-doped sol-gel films on Si as function of annealing temperature. Inset: PL peaks at 1.067 eV *vs.* 0.807 eV (dotted lines order annealing temperatures, increasing clockwise).

Variation of the peak Si emission intensity with anneal temperature is presented in Fig. 2, illustrating that emission is strongest for $T_a \approx 700$ °C. The PL is appreciable (up to 50 % of the maximum) for 600 – 800 °C and becomes rather weak for $T_a < 600$ °C (PL increases 5% from air-dried to $T_a = 500$-°C); PL is again low for $T_a > 900$ °C. The inset to Fig. 2 is a plot of the peak PL intensities at ~1.067 eV vs. those at ~0.807 eV; the implicit variable $T_a$ increases in clockwise rotation, 500 °C being at the lower left ($Er^{+3}$ PL for air-dried films is in the noise and therefore excluded). Linear fit to data logarithms, not shown, yields slope of 0.10 ± 0.17, i.e. near zero, and coefficient of linear correlation 0.9381. For the five samples yielding substantial ~1.067-eV emission (with emission levels varying by less than a factor of 2) the $Er^{+3}$ emission increases from near weakest to near strongest (with values varying by a factor of 600). The strength of the emission from the Si therefore does not depend on that from the $Er^{+3}$ ions in the silica film, from which we conclude that emissions at these two wavelengths arise from independent mechanisms. Emission at both wavelengths display a maximum in the dependence on $T_a$ (700 °C for the 1.067-eV peak, 850 °C for the 0.807-eV peak), hence the data points in the inset of Fig. 2 execute a cyclical pattern.

The observed annealing behavior points to enhancement of emission from the silicon through its interaction with the sol-gel film. Annealing sol-gel films for $T_a \sim 500$ °C removes virtually all of the solvents, organic compounds, and water, but leaves residual hydroxyls, which are detected by IR absorption signatures [18,19]. Films annealed at higher $T_a$ become denser, hydroxyl content decreases, and porosity is reduced [16,20,21]. In Fig. 3 we show the IR transmittance of five samples for 500 °C ≤ $T_a$ ≤ 900 °C (these samples were in the ambient for an extended period prior to measurement). The cut-off in transmittance below 1000 nm is from free carrier excitation in the Si substrate. Transmittance above 1200 is determined by film thickness and refractive index, as well as variability in back-side Si reflectance among the samples, owing to the use of single-side polished wafers. Structure is seen in the regions 1360 – 1480 nm and 1650 – 1760, which are close to where absorption from overtone vibrations of adsorbed water is observed in annealed monolithic silica gels [22]; the transmittance dip at 1720 nm (identified with overtone OH stretching vibrations in H-bonded H$_2$O [21]) is most pronounced for the sample annealed at 700 °C.

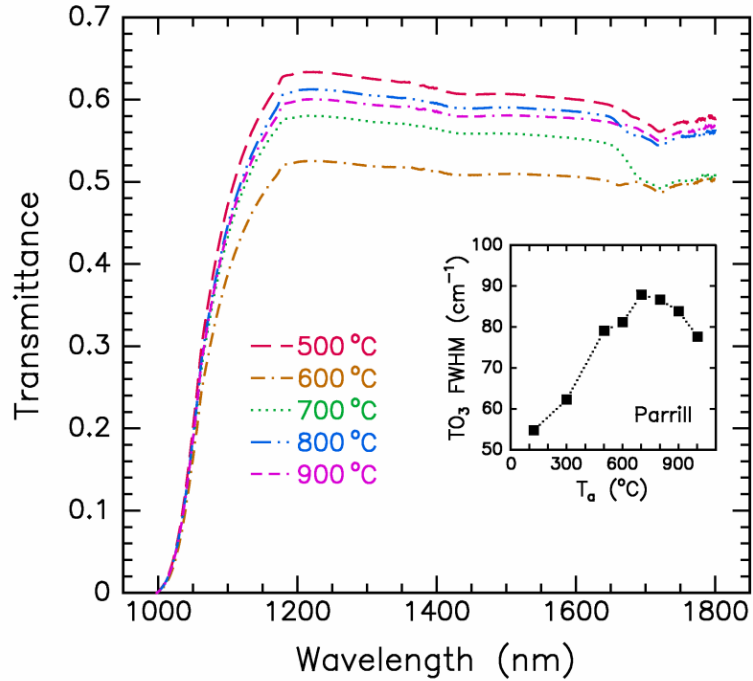

**Figure 3.** Infrared transmittance for samples of Er-doped sol-gel films on Si annealed at temperatures indicated. Inset: full-width at half-maximum of TO$_3$ vibrations in undoped sol-gel silica films on Si (from [20]).

Accepting that the near band-gap emission originates from a modification of the Si substrate, perturbation in the semiconductor bands in some form is worth bearing in mind. Formation of shallow p-n junctions by surface diffusion, as in the dopant diffusion method of [14], appears improbable in this case, since high-temperature diffusion from an Er-doped sol-gel film has been determined to be insufficient for doping Si with Er [23], owing to low solid solubility and diffusivity of Er in Si (note: the sol-gel films contain only Er as a potential dopant to underlying substrate). The results also indicate that the PL is not caused by damage in the silicon, such as produced in the ion implant method [11,12]. Once formed, extended topological defects, dislocations and stacking faults are rather stable for annealing up to 950 °C, rendering any PL signal that might associated with them to persist rather than to follow the observed trend of annealing away (Fig. 2).

Inherently reversible perturbations in the Si substrate, on the other hand, can be induced by oxide defects and stress. Pure-silica sol-gel films are known to be electrically leakier and have higher densities of interface states than their thermally-grown counterparts, and these attributes of electronic quality improve with increasing annealing temperature [24,25]. Annealing in air yields better oxides than annealing in vacuum or dry $N_2$, which could simply be passivation of the Si surface by thermally grown oxide. Although reductions in oxide and interface defects could account for decreased PL observed for $T_a > 800$ °C, it is difficult to apply similar reasoning to $T_a < 600$ °C, since sol-gel films are comparatively more defective at lower $T_a$.

Stresses in sol-gel films, caused by shrinkage and porosity, are the more likely substrate interactions to consider, especially since silica-gel is known to have inhomogeneity stemming from the fractal polymerization of $SiO_2$ in sol-gel hydrolysis [16,21]. Thermal densification produces large stresses [21], particularly for annealing above 400 °C, where mean tensile stress on the order of ~ 0.5 GPa was determined for sol-gel films on Si wafers [20]. Film stresses reduce the frequency and increase the width of IR absorption from asymmetric Si-O-Si stretching vibrations ($TO_3$ modes, 1060 – 1080 $cm^{-1}$) [21], which are interpreted as decreased and more broadly distributed tetrahedral bond angles. In a prior study of spin-coated sol-gel films on Si (similar to the present process, except for the Er doping) the lowest frequency and greatest broadening of the $TO_3$ mode occurred for annealing at $T_a = 700$ °C [20] (IR line width data reproduced in Fig. 3, inset), which coincidentally is where the strongest Si PL from our samples is observed. These IR data also indicate that bond-angle distortion decreases at high temperatures ($T_a \geq 900$ °C), where glassy reflow allows stress relief in $SiO_2$ films. Given that Si PL is comparatively subdued for $T_a < 600$ °C as well as for $T_a > 900$ °C, one may conclude that enhanced PL at $T_a \sim 700$ °C is to be associated with non-uniformities in film stresses, rather than with large mean film stress alone.

Inhomogeneous film stresses, acting in concert with native interfacial roughness, is therefore expected to create random elastic strains in the Si, produce random shifts in electron and hole bands, and modulate the Si band gap. A similar scenario of band gap fluctuations was previously invoked to explain enhanced Si emission from implanted (and presumably also diffused) junctions, by virtue of inhomogeneities ("doping spikes") at dopant (B or P) concentrations near the solid solubility limit [13]. According to this mechanism free carriers are released by excitons trapped at defects owing to energy band distortions. (Alternative explanations have been proffered, such large strains near dislocations [11] and gettering effects of defects formed by the implant-anneal process [12]; however, these do not account for enhanced emission observed in dislocation-free diffused junctions [14].) Thus it is reasonable to surmise that the random strains in Si adjacent to sol-gel films produce localized band fluctuations, analogous to those from doping spikes, and are sufficient to increase the effective cross section for radiative free-carrier recombination. Applying this interpretation, enhanced emission is expected to be produced in only a thin layer at the Si surface (on the order of the film thickness, ~0.1 μm, or less); since the excitation radiation penetrates ~1 μm, quantum efficiency (QE) could actually be increased ~$10^3$-fold (i.e. ~10× larger than the observed enhancement), suggesting QE >1% (assuming QE ≤ 0.01% for CZ silicon [1,2]).

The position of the PL peak (1.067 eV) is at a slightly lower energy than that observed for implanted Si (1.081 eV) [12] or in mesoscopic oxidized porous Si (1.088 eV) [6], which suggests a slightly smaller average band gap. More closely related to the present study are structures comprising thin islanded Ge films grown epitaxially on Si as well as on thin Si-Ge alloy buffer layers; both types yield room temperature PL from the Si substrate (~1.065 eV peak), in addition to emissions at lower photon energies from the Ge islands [26]. Such films contain a strain pattern created by the Ge-Si lattice mismatch and islanded growth topography; the PL does not involve forming p-n junctions, as in this work.  Our data also show that the sample for $T_a$ = 700 °C yields both the strongest Si PL (Fig. 2) and the most pronounced IR absorption features (Fig. 3). This is consistent with IR signatures of hydration in sol-gel films (on Si), which have been correlated with high film stress [20].

In conclusion, near-band gap (1.067 eV) photoluminescence at room temperature has been observed in CZ Si wafer material by spin-coating and annealing sol-gel silica films doped with Er (emitting at 0.807 eV).  Emission from the Si, strongest for annealing at 700 °C, correlates with inhomogeneous stresses in sol-gel films (owing to strong 25% shrinkage), as indicated indirectly from IR absorption. Based on the independent annealing behavior of the two PL signals and comparison with prior studies of sol-gel film properties, it appears that the Er is not directly involved in the Si emission. The spin-coating process presented herein is notable for enhancing light emission from bulk-type Si (estimated QE ~1%) in the absence of any patterning or p-n junctions.

Partial support by Deanship of Academic Research at the University of Jordan, Project contract no. 1030 and Hamdi Mango Center for Scientific Research (HMCSR), the New Jersey Institute of Technology, and the U.S. National Renewable Energy Laboratory, and encouragement and support of N.M. Ravindra are gratefully acknowledged.  Publication on this work has appeared [27].